# Cosmology with Ultra-light Pseudo-Nambu-Goldstone Bosons

Joshua A. Frieman[1,2], Christopher T. Hill[3], Albert Stebbins[1], and Ioav Waga[1,4]

[1] *NASA/Fermilab Astrophysics Center*

*Fermi National Accelerator Laboratory, Batavia, IL 60510*

[2] *Department of Astronomy and Astrophysics, University of Chicago, Chicago, IL 60637*

[3] *Theoretical Physics Department*

*Fermi National Accelerator Laboratory, Batavia, IL 60510*

[4] *Universidade Federal do Rio de Janeiro, Instituto de Física*

*Rio de Janeiro, RJ, 21943 - Brasil*

(May 1995)

## Abstract

We explore the cosmological implications of an ultra-light pseudo-Nambu-Goldstone boson. With global spontaneous symmetry breaking scale $f \simeq 10^{18}$ GeV and explicit breaking scale comparable to MSW neutrino masses, $M \sim 10^{-3}$ eV, such a field, which acquires a mass $m_\phi \sim M^2/f \sim H_0$, would have become dynamical at recent epochs and currently dominate the energy density of the universe. The field acts as an effective cosmological constant for several expansion times and then relaxes into a condensate of coherent non-relativistic bosons. Such a model can reconcile dynamical estimates of the density parameter, $\Omega_m \sim 0.2$, with a spatially flat universe, and can yield an expansion age $H_0 t_0 \simeq 1$ while remaining consistent with limits from gravitational lens statistics.

Typeset using REVTEX

Recently, a cosmological model with substantial vacuum energy—a relic cosmological constant $\Lambda$—has come into vogue for several reasons. First, dynamical estimates of the mass density on the scales of galaxy clusters, the largest gravitationally bound systems, suggest that $\Omega_m = 0.2 \pm 0.1$ for the matter ($m$) which clusters gravitationally (where the density parameter $\Omega$ is the ratio of the mean mass density of the universe to the critical Einstein-de Sitter density, $\Omega(t) = 8\pi G\rho/3H^2$) [1]. However, if a sufficiently long epoch of inflation took place during the early universe, the present spatial curvature should be negligibly small, $\Omega_{tot} = 1$. A form of dark, homogeneously distributed energy density with $\Omega_h = 1 - \Omega_m$, such as a cosmological constant, is one way to resolve the discrepancy between $\Omega_m$ and $\Omega_{tot}$.

The second motivation for the revival of the cosmological constant is the 'age crisis' for spatially flat $\Omega_m = 1$ models. Current estimates of the Hubble expansion parameter from a variety of methods, most recently Cepheid variable stars in the Virgo cluster [2], are (with some notable exceptions) converging to relatively high values, $H_0 \simeq 80 \pm 15$ km/sec/Mpc [3], while estimates of the age of the universe from globular clusters are holding at $t_{gc} \simeq 13 - 15$ Gyr or more [4]. Thus, the 'expansion age' $H_0 t_0 = 1.14(H_0/80\text{km/sec/Mpc})(t_0/14\text{Gyr})$ is uncomfortably high compared to that for the standard Einstein-de Sitter model with $\Omega_m = 1$, for which $H_0 t_0 = 2/3$. On the other hand, for models with a cosmological constant, $H_0 t_0$ can be significantly larger: for example, for $\Omega_\Lambda \equiv \Lambda/3H_0^2 = 0.8 = 1 - \Omega_m$, one finds $H_0 t_0 = 1.076$. Third, cosmological constant-dominated models for large-scale structure formation with cold dark matter (CDM) and a nearly scale-invariant spectrum of primordial density perturbations (as predicted by inflation) provide a better fit to the observed power spectrum of galaxy clustering than does the 'standard' $\Omega_m = 1$ CDM model [5].

While they provide a number of theoretical benefits, models with a relic cosmological constant have problems of their own. A cosmological constant for which, e.g., $\Omega_\Lambda \sim 1$ corresponds to a vacuum energy density $\rho_{vac} = \Lambda/8\pi G \simeq (0.003 \text{ eV})^4$. Within the context of quantum field theory, there is as yet no understanding of why the vacuum energy density arising from zero–point fluctuations is not of order the Planck scale, $M_{Pl}^4$, or at least of order the supersymmetry breaking scale, $M_{SUSY}^4 \sim \text{TeV}^4$, both many orders of magnitude larger. Within the context of classical field theory, there is no understanding of why the vacuum energy density is not of the order of the scale of one of the vacuum condensates, such as $-M_{GUT}^4$, $-M_{SUSY}^4$, $-M_W^4 \sin^4\theta_W/(4\pi\alpha)^2 \sim (175 \text{ GeV})^4$, or $-f_\pi^4 \sim (100 \text{ MeV})^4$. Thus, a vacuum density of order $(0.003 \text{ eV})^4$ appears to require cancellation between two (or more) large numbers to very high precision. In addition, it implies that we are observing the universe just at the special epoch when $\Omega_m$ is comparable to $\Omega_\Lambda$, which might seem to beg for further explanation.

Moreover, such models now face strong observational constraints from gravitational lens statistics: in a spatially flat universe with non-zero $\Lambda$, the lensing optical depth at moderate redshift is substantially larger than in the Einstein-de Sitter model with $\Omega_m = 1$ [6]. In the Hubble Space Telescope Snapshot Survey for lensed quasars, there are only four lens candidates (thought to be lensed by foreground galaxies) in a sample of 502 QSOs; from this data, the bound $\Omega_\Lambda \lesssim 0.6 - 0.8$ has been inferred [7]. For $\Omega_\Lambda = 1 - \Omega_0 < 0.7$, the expansion age satisfies $H_0 t_0 < 0.96$. With a cosmological constant saturating this bound, the globular cluster age $t_0 \geq 14$ Gyr implies $H_0 < 67$ km/sec/Mpc, within the uncertainties of but below the central value of recent Hubble parameter determinations.

It is conventional to assume that the fundamental vacuum energy of the universe is zero, owing to some as yet not understood mechanism, and that this new physical mechanism 'commutes' with other dynamical effects that lead to sources of energy density (after all, there is gravitational energy density acting on cosmological scales). This is required so that, e.g., at earlier epochs there can temporarily exist non-zero vacuum energy which allows inflation to take place, but the situation in reality could be more complex. Nonetheless, if this simple hypothesis is the case, then the effective vacuum energy at any epoch will be dominated by the heaviest fields which have not yet relaxed to their vacuum state. At late times, these fields must be very light. This is a big asumption: the cosmological 'constant' may be in the process of relaxing in a self-consistent way which leaves a residual effect at any scale, and we can only hope that this hypothesis approximates this possibility.

Adopting this working hypothesis, in this Letter we explore the consequences of an ultra-light pseudo-Nambu–Goldstone boson (hereafter, PNGB) field which is (i) *currently* relaxing to its vacuum state and which (ii) dynamically dominates the energy density during the epoch in which it relaxes. PNGB models are characterized by two mass scales, a spontaneous and an explicit symmmetry breaking scale; we will see that the two dynamical conditions above essentially fix these two mass scales to values which are 'reasonable' from the viewpoint of particle physics. Since these scales can have a plausible origin in particle physics models, we may have an explanation for the 'coincidence' that the vacuum energy is dynamically important at the present epoch. Moreover, in these models, the cosmological constant is evanescent, within a few expansion times converting into scalar field oscillations which subsequently redshift as non-relativistic matter. Thus, unlike cosmological constant-dominated models, the universe is not now entering a phase of exponential de Sitter expansion, but has rather undergone a brief hiatus of quasi-accelerated expansion. As a byproduct, we shall see that the gravitational lens constraints on $H_0 t_0$ in this model are slightly less severe than for cosmological constant models, allowing an expansion age as large as $H_0 t_0 \simeq 1.05$.

In particle physics, the best known example of a PNGB is the ordinary $\pi$ meson (the longitudinal $W$ and $Z$ bosons are actually exact Nambu–Goldstone bosons in association with gauge fields). An example of a very light hypothetical PNGB is the axion, associated with the Peccei-Quinn symmetry introduced to solve the strong CP problem [8]. Axions arise when a global $U(1)_{PQ}$ symmetry is spontaneously broken by the vacuum expectation value of a complex scalar at the scale $f_a$, $\langle \Phi \rangle = f_a e^{ia/f_a}$; at this scale, the axion, the angular field $a$ around the infinitely degenerate minimum of the potential, is a massless Nambu-Goldstone boson. QCD instantons explicitly break the global symmetry at the scale $f_\pi \sim 100$ MeV, generating the axion mass, $m_a \sim O(m_\pi f_\pi / f_a)$. Since its couplings and mass are suppressed by inverse powers of $f_a$, the axion is very light and very weakly interacting. Nevertheless, it can play an important role in astrophysics and cosmology; indeed, astrophysical and cosmological arguments constrain the global symmetry breaking scale to lie in a narrow window around $f_a \sim 10^{10} - 10^{12}$ GeV. Thus, the axion mass $m_a \sim 10^{-5} \text{eV}(10^{12}\text{GeV}/f_a)$, and its Compton wavelength is macroscopic, $\lambda_a \sim (f_a/10^{12}\text{GeV})$ cm.

Although motivated by the strong CP problem, the axion is a particular instance of a more general phenomenon that includes familons, majorons, [9] and more exotic objects [10]. In all these models, the key ingredients are the scale of spontaneous symmetry breaking $f$ (at which the effective Lagrangian still retains the symmetry) and a scale of explicit symmetry breaking $\mu$ (at which the effective Lagrangian contains the explicit symmetry breaking term). The mass of the PNGB is then $m_\phi \sim \mu^2/f$. Ref. [11] introduced a class of PNGBs closely related to familons (called 'schizons'), with masses $m_\phi \simeq m_{fermion}^2/f$. Models in which $m_{fermion}$ is associated with a hypothetical neutrino mass, $m_\nu \sim 0.001 - 0.01$ eV, and $f \sim M_{GUT} - M_{Pl} \sim 10^{15} - 10^{19}$ GeV, were studied in ref. [12] in the context of late time phase transitions [13] and form the theoretical basis for the present work. In this case, the PNGB Compton wavelength $m_\phi^{-1}$ is comparable to cosmological distance scales.

From the viewpoint of quantum field theory, pseudo-Nambu-Goldstone bosons are the only way to have naturally ultra–low mass, spin–0 particles. In this regard, 'technically' natural small mass scales are those which are protected by symmetries, such that when the small masses are set to zero, they cannot be generated in any order of perturbation theory, owing to the restrictive symmetry. For generic PNGBs, when the symmetry breaking scale $\mu$ is set to zero, the symmetry becomes exact, and radiative corrections do not yield an explicit symmetry breaking term (the radiative corrections are "multiplicative" of the scale $\mu$ in this situation). In the ultra-light PNGB models mentioned above, the small mass $m_\phi$ is protected by fermionic chiral symmetries (and additional discrete symmetries) and is therefore technically natural. That is, when certain fermion mass terms are set to zero in the Lagrangian, the PNGB mass goes to zero; the fermion mass terms will not be generated in any order of perturbation theory.

As an example, consider the $Z_N$-invariant low-energy effective chiral Lagrangian for $N$

neutrinos [12],

$$\mathcal{L} = \frac{1}{2}\partial_\mu\phi\partial^\mu\phi + \sum_{j=0}^{N-1} \bar{\nu}_j i\gamma^\mu \partial_\mu \nu_j + \left(m_0 + \epsilon e^{i(\phi/f + 2\pi j/N)}\right)\bar{\nu}_{jL}\nu_{jR} + h.c. \qquad (1)$$

where $\nu_{(R,L)}$ are respectively right– and left–handed projections, $\nu_{(R,L)} = (1 \pm \gamma^5)\nu/2$. The term proportional to $\epsilon$ can arise from a Yukawa coupling $g\bar{\nu}_L \nu_R \Phi + h.c.$, where the complex scalar field $\Phi$ has a non-zero vacuum expectation value, $\langle\Phi\rangle = f e^{i\phi/f}/\sqrt{2}$, and $\epsilon \equiv gf/\sqrt{2}$. The term proportional to $m_0$ is an explicit breaking which usually comes from some deeper breaking in the theory. In the limit $m_0 \to 0$, this is a familiar chiral Lagrangian, possessing a continuous $U(1)$ chiral symmetry. The $U(1)$ chiral symmetry is broken to a residual $Z_N$ discrete symmetry:

$$\nu_j \to \nu_{j+1}\,;\quad \nu_{N-1} \to \nu_0\,;\quad \phi \to \phi + 2\pi j f/N\,. \qquad (2)$$

The induced one-loop correction, with cutoff $\Lambda < f$, is

$$\mathcal{L}_{1-loop} = \sum_{j=0}^{N-1} \frac{M_j^4}{16\pi^2} \ln\left(\frac{\Lambda^2}{M_j^2}\right)\,, \qquad (3)$$

where

$$M_j^2 = m_0^2 + \epsilon^2 + 2m_0\epsilon \cos\left(\frac{\phi}{f} + \frac{2\pi j}{N}\right)\,, \qquad (4)$$

which respects the discrete symmetry. For $N = 2$, the leading contribution is log divergent, and the induced PNGB mass is of order $m_\phi \sim m_0\epsilon/f$; if $\epsilon \sim m_0 \sim m_\nu$, then $m_\phi \sim m_\nu^2/f$. For $N > 2$, the sum $\Sigma_j M_j^4$ is independent of $\phi$; thus, the $\phi$-dependent term is independent of the cutoff $\Lambda$, and for $N > 2$ we can write the 1-loop effective potential,

$$V(\phi) = -\sum_j \frac{M_j^4}{16\pi^2} \ln M_j^2\,. \qquad (5)$$

In this case, the $\phi$-potential is explicitly calculable, and one again finds a quasi-periodic potential with mass scale $m_\phi \sim m_\nu^2/f$.

We are thus led to study the cosmological evolution of a light scalar field $\phi$ with effective Lagrangian

$$\mathcal{L} = \frac{1}{2}\partial_\mu\phi\partial^\mu\phi - M^4[\cos(\phi/f) + 1]\,. \qquad (6)$$

The theory is determined by two mass scales, $M$, which from (1) is expected to be within an order of magnitude of a light fermion (neutrino) mass, and $f$, the global symmetry breaking scale. Since $\phi$ will turn out to be extremely light, we assume that it is the only classical field which has not yet reached its vacuum expectation value. Thus, in accordance with our working hypothesis, the constant term in the PNGB potential has been chosen to ensure that the vacuum energy vanishes at the minimum of the $\phi$ potential. We focus upon the spatially homogeneous, zero-momentum mode of the field, $\phi(t) = \langle \phi(\vec{x},t) \rangle$, where the brackets denote spatial averaging. We are assuming that the spatial fluctuation amplitude $\delta\phi(\vec{x},t)$ is small compared to $\phi(t)$, as would be expected after inflation if the post-inflation reheat temperature $T_{RH} < f$: in this case, aside from inflation-induced quantum fluctuations (which correspond to isocurvature density perturbations [15]), the field will be homogeneous over many present Hubble volumes. Since we will be interested in the case $f \sim M_{Pl}$ (see below), this is not a significant restriction. Finally, for simplicity we assume that any finite-temperature corrections to the potential $V(\phi)$ in (6) are unimportant at the epochs of interest (this is different from the case of axions, for which finite-temperature corrections do affect the axion field evolution). The scalar equation of motion is then

$$\ddot{\phi} + 3H\dot{\phi} + dV(\phi)/d\phi = 0 , \qquad (7)$$

where the Hubble parameter is given by $H^2 = (\dot{a}/a)^2 = (8\pi/3M_{Pl}^2)(\rho_m + \rho_\phi)$ for a spatially flat universe, $\Omega_m + \Omega_\phi = 1$, $a(t)$ is the cosmic scale factor, and $\Omega_m$ is the density parameter of non-relativistic matter (e.g., baryons and/or weakly interacting massive particles). We will focus on recent epochs, when the radiation energy density is negligible compared to non-relativistic matter.

The cosmic evolution of $\phi$ is essentially determined by the ratio of its mass, $m_\phi \sim M^2/f$, to the instantaneous expansion rate, $H(t)$. For $m_\phi \lesssim 3H$, the field evolution is overdamped by the expansion, and the field is effectively frozen to its initial value. Since $\phi$ is initially laid down in the early universe (at a temperature $T \sim f \gg M$) when its potential was dynamically irrelevant, its initial value in a given Hubble volume will generally be displaced from its vacuum expectation value $\phi_m = \pi f$ (vacuum misalignment). Thus, at early times, the field acts as an effective cosmological constant, with vacuum energy density and pressure $\rho_\phi \simeq -p_\phi \sim M^4$. At late times, $m_\phi \gg 3H(t)$, the field undergoes damped oscillations about the potential minimum; at sufficiently late times, these oscillations are approximately harmonic, and the stress-energy tensor of $\phi$ averaged over an oscillation period is that of non-relativistic matter, with energy density $\rho_\phi \sim a^{-3}$ and pressure $p_\phi \simeq 0$.

Let $t_x$ denote the epoch when the field becomes dynamical, $m_\phi = 3H(t_x)$, with corresponding redshift $1 + z_x = (a(t_0)/a(t_x)) = (M^2/3H_0 f)^{2/3}$; for comparison, the universe makes the transition from radiation- to matter-domination at $z_{eq} \simeq 2.3 \times 10^4 \Omega_m h^2$ [where $h = H_0/(100 \text{ km/sec/Mpc})$]. The $f - M$ parameter space is shown in Fig. 1. To the right of the diagonal line $m_\phi = 3H_0$, the field becomes dynamical before the present epoch and currently redshifts like non-relativistic matter; to the left of this line, $\phi$ is still frozen and currently acts like a cosmological constant (the region denoted by '$\Lambda$'). In the dynamical region, the present density parameter for the scalar field is approximately $\Omega_\phi \simeq 24\pi(f/M_{Pl})^2$, independent of $M$ [12] (assuming the initial field value $\phi_i = \mathcal{O}(1)f$); thus, the horizontal line at $f = 1.4 \times 10^{18}$ GeV indicates the cosmic density limit $\Omega_\phi = 1$. In the frozen ($\Lambda$) region, on the other hand, $\Omega_\phi$ is determined by $M^4$, independent of $f$, and the bound $\Omega_\phi = 1$ is indicated by the vertical line.

Focus on the dynamical region in the right-hand portion of Fig. 1. If $\phi$ dominates the energy density of the Universe, the growth of density perturbations is strongly suppressed for physical wavenumbers larger than the 'Jeans scale' [16] $k_J \simeq m_\phi(\phi_m(t)/M_{Pl})^{1/2}$, where $\phi_m(t) \sim f[(1 + z(t))/(1 + z_x)]^{3/2}$ is the amplitude of the homogeneous field oscillations at $z(t) < z_{eq}$. If this Jeans scale is too large, perturbations on galaxy and cluster scales would not grow at high redshift, leading to a power spectrum with an unacceptably large coherence scale. We can express the resulting perturbation power spectrum in terms of the standard cold dark matter (CDM) spectrum as $P(k) = P_{cdm}(k)F^2(k)$; for $z_x > z_{eq}$, the relative suppression factor due to the scalar field is [17]

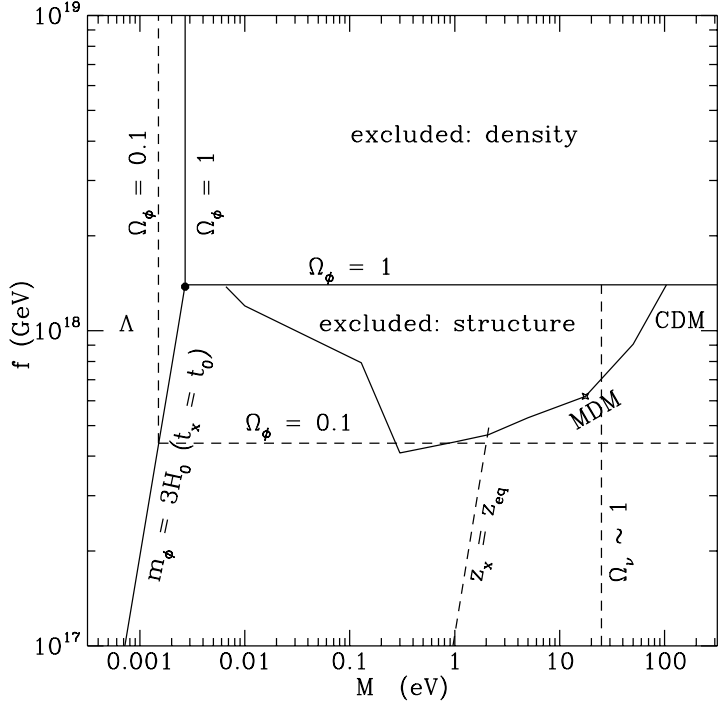

Fig. 1: The PNGB model parameter space.

$$F(k) \simeq \left(\frac{1+z_{eq}}{1+z_*(k)}\right)^{(5/4)[(1-24\Omega_\phi/25)^{1/2}-1]}$$
$$= \left[\left(\frac{110\ h\ \text{eV}}{M}\right)\left(\frac{k}{1h\text{Mpc}^{-1}}\right)\right]^{5[(1-72.4(f/M_{Pl})^2)^{1/2}-1]}$$

Here, $1+z_*(k) = [(M/k)(3H_0/M_{Pl})^{1/2}]^4$ is the redshift at which the physical wavenumber $k_{phys} = k(1+z)$ drops below $k_J$, so that scalar perturbations on that scale can begin to grow. Thus, $M$ sets the scale where the power spectrum turns down from the CDM spectrum, and $f$ (through $\Omega_\phi$) determines the spectral slope $n$ of the suppression factor, $F(k) \sim k^{-n}$ with $-4 \leq n \leq 0$ (note that for $\Omega_\phi \lesssim 0.2$, $n \simeq 12\Omega_\phi/5$). For galaxies and quasars to form at moderate redshift, the power at small scales should not be very strongly suppressed compared to standard CDM. We therefore impose the appoximate bound $F(k = 1.6h\text{Mpc}^{-1}) > 0.3$, which corresponds to the curved boundary in Fig. 1: the region above this curve is excluded. To the right of this region (in the area marked CDM), $\phi$ acts as an ordinary cold dark matter candidate, a lighter version of the dark matter axion. In the area marked MDM, the effects of $\phi$ on the small-scale power spectrum are similar to those of a light neutrino in the mixed dark matter model: at the point marked by the star, the variance of the density field smoothed with a top-hat window of radius $R = 8h^{-1}$ Mpc is $\sigma_8(\phi) \simeq \sigma_8^{cdm}/2$. When the amplitude is normalized to COBE on large scales, this yields $\sigma_8(\phi) \simeq 0.6$, as suggested by the abundance of rich clusters of galaxies and the small-scale pairwise velocity dispersion of galaxies. In this region of parameter space, the neutrinos of mass $m_\nu \sim M \sim$ several eV could play a dynamical role in structure formation as well.

For the remainder of this Letter, we focus on the parameter region near the bullet in Fig. 1, in which the field becomes dynamical at recent epochs, $z_x \sim 0 - 3$, or in the near future: this has new consequences for the classical cosmological tests and the expansion age, and it does not lead to the small-scale power suppression above. We thus impose the constraint $m_\phi = M^2/f \lesssim 3H_0$. The second condition is that the PNGB energy density be dynamically relevant for the recent expansion of the universe, which implies $\rho_\phi(t_0) \sim \rho_{crit}(t_0)$, or $M^4 \simeq 3H_0^2 M_{Pl}^2/8\pi$. Combining these two constraints determines the two mass scales in the theory to be $f \gtrsim M_{Pl}/(24\pi)^{1/2} \simeq 10^{18}$ GeV and $M \simeq 3 \times 10^{-3} h^{1/2}$ eV. As argued above, we can construct particle physics models for light PNGBs with these mass scales: the spontaneous breaking scale $f$ is comparable to the Planck scale, and the explicit breaking scale $M$ is comparable to that expected for light neutrinos for the MSW solution to the solar neutrino problem. The mass of the resulting PNGB field is miniscule, $m_\phi \lesssim 4 \times 10^{-33}$ eV, and (by construction) its Compton wavelength is of order the current Hubble radius, $\lambda_\phi = m_\phi^{-1} = H_0^{-1}/3 \gtrsim 1000 h^{-1}$ Mpc [18]. This is a generic feature of scalar field models for relic vacuum energy that satisfy $V(\phi_m) = 0$.

Figure 2 shows several examples of the evolution of the scalar field [Eqn.(7) with the potential of Eqn.(6) and the Hubble parameter given by the expression immediately below Eqn.(7)]. We show $\Omega_m = 1 - \Omega_\phi$ as a function of the expansion age $Ht$, for different initial values of the field $\phi_i/f$ (assuming $\dot{\phi}_i = 0$, since the field is Hubble-damped at early times). The numerical evolution starts at $\rho_m/M^4 \gg 1$, *i.e.*, at the top of the figure ($\Omega_m \simeq 1 \gg \Omega_\phi$) in the matter-dominated epoch. At early times, the field is effectively frozen to its initial value by the Hubble damping term in Eqn.(7), and the evolution tracks that of a cosmological constant model (curve labelled 'vac' in Fig. 2). At $t \sim t_x$, the field begins to roll classically; on a timescale initially comparable to the expansion time, the expansion age $Ht$ reaches a maximum and subsequently falls toward 2/3 (indicated with the vertical dashed line) as the field undergoes Hubble-damped oscillations about the potential minimum. The evolutionary tracks are universal: a shift in the mass scale $f$ accompanied by an appropriate rescaling of the initial field value $\phi_i$ leads to essentially identical tracks, *i.e.*, a given track actually corresponds to a family of choices of $(\phi_i, f)$.

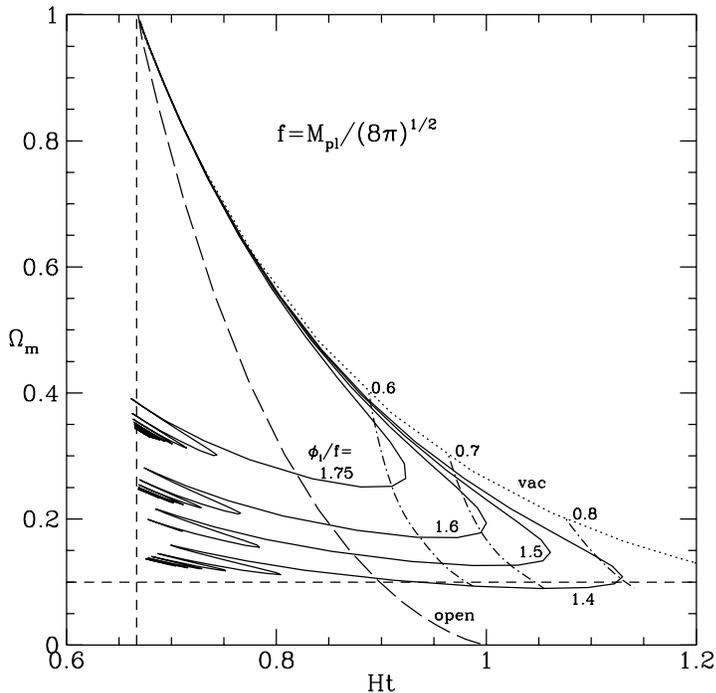

Fig. 2: The non-relativistic mass density $\Omega_m = 1 - \Omega_\phi$ vs. $Ht$, for $f = M_{Pl}/\sqrt{8\pi}$. The solid curves correspond to several initial values for the field, $\phi_i/f = 1.4, 1.5, 1.6,$ and 1.75. The evolution starts at the top of the figure and ends at the lower left. The vertical dashed line shows the Einstein-de Sitter expansion age $Ht = 2/3$, the horizontal dashed line shows the lower bound $\Omega_m = 0.1$ from dynamical mass estimates, the dotted curve (labelled 'vac') shows the evolution for a cosmological constant model, and the long-dashed curve corresponds to an open model with $\Omega_\phi = 0$. The dot-dashed curves (labelled 0.6, 0.7, 0.8) bracket the constraints from lensed QSOs in the HST snapshot survey (see text).

The observational consequences of this model follow when one identifies the present epoch

$t_0$ on an evolutionary track—this implicitly corresponds to fixing the mass scale $M$. For a given expansion age $H_0 t_0$, one can choose the upper branch, where the field is still frozen and thus nearly identical to a cosmological constant, or the lower (dynamical) branch, for which the recent evolution will be intermediate between vacuum- and matter-dominated and which has qualitatively new features. Dynamical estimates of the mass in galaxy clusters indicate the lower bound $\Omega_m \gtrsim 0.1$ for the mass density in non-relativistic matter. Consequently, the lower branch is excluded if the initial value of the field is below some value, *e.g.*, $\phi_i/f \simeq 1.3$ for $f = M_{Pl}/\sqrt{8\pi}$. Physically, for such small values of $\phi_i/f$, the universe undergoes several e-foldings of inflation before the field begins to oscillate, diluting the density of non-relativistic matter. Consequently, to achieve large expansion times in this model, $H_0 t_0 \sim 1$, the present epoch must be in the vicinity of the 'nose' of the evolutionary track, which corresponds approximately to the condition $t_x \sim t_0$ imposed above.

As with vacuum-dominated models, these scalar field models can in principle reach arbitrarily long expansion ages, $Ht \gg 1$, if $\phi_i/f$ is sufficiently small. However, this region of parameter space is excluded by the observed statistics of gravitationally lensed quasars. The 3 dot-dashed curves in Fig. 2 show the observed constraints on the incidence of lensed QSOs. We computed the number of lensed QSOs expected in the HST Snapshot survey [19] for cosmological constant models with $\Omega_\Lambda = 0.6$, 0.7, and 0.8; along the 3 curves in Fig. 2, the number of expected lensed QSOs in the PNGB models are equal to these 3 values. Since different assumptions about galaxy models yield different lensing fractions, we show the limits corresponding to these three cases to cover the spread of quoted limits in the literature [7] (the region to the right of each curve is excluded). For a given lensing limit, the upper bound on the expansion age $H_0 t_0$ is increased in the scalar field models compared to the cosmological constant model; imposing the lower bound $\Omega_m > 0.1$, the bound on $H_0 t_0$ can be relaxed by $7 - 10\%$. Thus, the scalar field models are relatively more successful than a cosmological constant at easing the 'age crisis' while remaining within the observational constraints, provided $\Omega_m$ is fairly low.

We have presented a class of models which give rise to (technically natural) ultra-light pseudo-Nambu-Goldstone bosons. With spontaneous and explicit symmetry breaking scales comparable to those plausibly expected in particle physics models, the resulting PNGB becomes dynamical at recent epochs and currently dominates the energy density of the universe. Such a field acts as a form of smoothly distributed dark matter, with a stress tensor at the current epoch intermediate between that of the vacuum and non-relativistic matter. Such a model 'explains' the coincidence between matter and vacuum energy density in terms of particle physics mass scales, reconciles low dynamical mass estimates of the density parameter, $\Omega_m \sim 0.2$, with a spatially flat universe, and does somewhat better than a cosmological constant at alleviating the 'age crisis' for spatially flat cosmologies while remaining within the observational bounds imposed by gravitational lens statistics.

This work was supported by the DOE and NASA grant NAG5-2788 at Fermilab. JF thanks Lloyd Knox for his scalar field evolution code and Richard Watkins for discussions. After this work was completed, we became aware of related work by Fukugita and Yanagida [20], which considers an axion model for the non-dynamical ($\Lambda$) region of Fig. 1.

# REFERENCES

[1] Dynamical estimates of the density parameter on larger scales using peculiar velocities have so far been inconclusive, with estimates falling in the range $\Omega_m \sim 0.2 - 1$. Cf. A. Dekel, ARAA **32**, 371 (1994); S. Cole, K. Fisher, and D. Weinberg, preprint IASSNS-AST 94/63 (1994), submitted to MNRAS.

[2] W. Freedman, etal., Nature **371**, 757 (1994). M. Pierce, etal., Nature **371**, 385 (1994).

[3] G. Jacoby, etal., PASP **104**, 559 (1992); J. Tonry, Ap.J.Lett. **373**, L1 (1991).

[4] A. Renzini, in *Proc. 16th Texas Symposium on Relativistic Astrophysics and 3rd Symposium on Particles, Strings, and Cosmology*, eds. C. Akerlof and M. Srednicki, (New York Academy of Sciences, New York, 1992); X. Shi, Ap.J., in press (1995).

[5] G. Efstathiou, S. Maddox, and W. Sutherland, Nature **348**, 705 (1990); L. Kofman, N. Gnedin, and N. Bahcall, Ap.J.**413**, 1 (1993); J. Peacock and S. Dodds, MNRAS **267**, 1020 (1994).

[6] M. Fukugita and E. L. Turner, MNRAS **253**, 99 (1991).

[7] C. Kochanek, Ap.J. **419**, 12 (1993); D. Maoz and H.-W. Rix, Ap.J. **416**, 425 (1993).

[8] R. Peccei and H. Quinn, Phys. Rev. Lett. **38**, 1440 (1977); S. Weinberg, Phys. Rev. Lett. **40**, 223 (1978); F. Wilczek, Phys. Rev. Lett. **46**, 279 (1978); for a review, see J. Kim, Phys. Rep. **150**, 1 (1987). For reviews of astrophysical constraints on axions, see M. S. Turner, Phys. Rep. **197**, 67 (1990) and G. Raffelt, Phys. Rep. **198**, 1 (1990).

[9] F. Wilczek, Phys. Rev. Lett. **49**, 1549 (1982); G. Gelmini, S. Nussinov, and T. Yanagida, Nucl. Phys. **B219**, 31 (1983); Y. Chigashige, R. Mohopatra, and R. Peccei, Phys. Lett. **B241**, 96 (1990).

[10] P. Vorobev and Y. Gitarts, Phys. Lett. **B208**, 146 (1988); A. Anselm, Phys. Rev. **D37**, 2001 (1990); A. Anselm and N. Uraltsev, Phys. Lett. **B116**, 161 (1982)

[11] C. T. Hill and G. G. Ross, Nucl. Phys. **B311**, 253 (1988); Phys. Lett. **B203**, 125 (1988).

[12] J. Frieman, C. Hill, and R. Watkins, Phys. Rev. **D46**, 1226 (1992). Another possibility for an ultralight PNGB (discussed in this reference) is an axion that couples to a (presumably hidden) gauge group that becomes strong at the scale $\Lambda \sim 10^{-3}$ eV.

[13] C. T. Hill, D. N. Schramm, and J. Fry, Comm. Nucl. Part. Phys. **19**, 25 (1989); I. Wasserman, Phys. Rev. Lett. **57**, 2234 (1986); W. H. Press, B. Ryden, and D. Spergel, Ap. J. **347**, 590 (1989).

[14] G. 't Hooft, in *Recent Developments in Gauge Theories*, eds. G. 't Hooft, *et al*, (Plenum Press, New York and London, 1979), p. 135.

[15] The isocurvature density fluctuation amplitude at Hubble-radius crossing is approximately $\delta\rho/\rho \sim H_i/\phi_i \sim H_i/f$, where $H_i$ is the Hubble parameter during inflation and $\phi_i \sim f$ is the initial misalignment of the field in our Hubble volume. Constraints from the microwave background anisotropy on the amplitude of adiabatic perturbations in inflation typically yield the upper bound $H_i \lesssim 10^{13}$ GeV, so the amplitude of isocurvature perturbations is sufficiently small if $f \sim 10^{18}$ GeV.

[16] M. Khlopov, B. Malomed, and Y. Zel'dovich, MNRAS **215**, 575 (1985).

[17] By construction, this expression applies where $F(k) \leq 1$. If $z_x < z_{eq}$, $z_x$ replaces $z_{eq}$ in the numerator. If $1 + z_*(k) \leq 1$, it is replaced by unity in the denominator.

[18] B. Ratra and P. J. E. Peebles, Phys. Rev. **D37**, 3406 (1988), considered the cosmological consequences of a presently rolling scalar field, but they assumed ad hoc scalar potentials (generally exponentials or negative power laws) with rather different forms and consequences from the model here.

[19] D. Maoz, etal., Ap.J. **409**, 28 (1993).

[20] M. Fukugita and T. Yanagida, preprint YITP/K-1098 (1995).